\documentclass[12 pt]{article}
\usepackage{amsmath, amsfonts, amssymb}
\usepackage{graphicx}
\usepackage{indentfirst}
\usepackage{threeparttable}
\usepackage{url}

\begin{document}

\title{\LARGE \bf Analysis of the magnetic fields in massive young stellar objects with masers}
\author{\bf 
$^{1}$P.~I.~Pavlova\thanks{E-mail: pavlovapolina902683@gmail.com}, $^{2,1}$S.~A.~Khaibrakhmanov
and $^{1}$A.~M.~Sobolev}
\date{\it  \small  $^1$ Ural Federal University , \\
19 Mira street, 620062 Ekaterinburg, Russia \\
$^2$ Saint-Petersburg State University , \\
7-9 Universitetskaya Embankment, St. Petersburg, Russia, 199034}

\maketitle

\begin{abstract}
We analyze observational data on methanol masers in the disks of young massive stellar objects in the massive star formation regions: NGC6334I, G33.641-0.228, G12.89+0.49. Special attention is paid to analysis of the magnetic fields and their possible connection with maser flares. For this purpose, we estimate the distance from star to maser, the magnetospheric radius and magnetic field strength in the region of this radius. The distances from stars to masing regions lie in range from $380$ to $620$~au. Stellar and disk's magnetic field strengths are comparable at the magnetopshere boundary and lie in the range from $590$ to $1880$~G for considered objects. Our estimates show that the magnetic fields may play an important role in the considered regions. In particular, observed luminosity bursts and maser flares may be related to magnetic reconnection events near the stars.
\end{abstract}


\section{Introduction}
Massive young stellar objects (MYSOs) are massive stars, $M>10\,M_\odot$, at an early stage of evolution, deeply embedded in their envelopes. The MYSOs have a similar nature and structure to young stellar objects in the regions of low-mass star formation.  These objects form as a result of the gravitational collapse of molecular cloud cores. The collapse ends up in the formation of a star surrounded by a circumstellar accretion disk. The disk contributes to the infrared excess of these young objects, while their luminosity is determined by a sum of stellar and accretion luminosities. In contrast to their low-mass counterparts, the MYSOs are much less studied due to their remoteness (Beuther et al., 2025; Ellingsen et al., 2007).

A characteristic feature of MYSOs in high-mass star formation regions (HMSFR) is the presence of masers (Ellingsen et al., 2005). A maser is a phenomenon of amplification of the intensity of radiation passing through a region where there is an inverted population of molecular energy levels (Elitzur, 1992; Cragg, 2002). The maser effect has been detected in the emission of such molecules as methanol, water, hydroxyl, etc. Masers are found in jets and outflows from MYSOs, circumstellar disks, boundaries of HII regions. Masers are divided into two main classes, depending on their pumping mechanisms (Menten, 1993). The class \MakeUppercase{\romannumeral1} masers are characterised by excitation resulting from collisions with surrounding particles. In contrast, class \MakeUppercase{\romannumeral2} masers are excited by interaction with radiation, which also contributes to population inversion. Under certain conditions, maser lines are prone to detectable Zeeman effect, which can be used to measure the magnetic field strength in the masing region (Vlemmings, 2008). 

Maser emission in HMSFR is often variable (Shakhvorostova et al., 2018; Hafner et al., 2024; Ashimbaeva et al., 2024). The variability can be periodic and stochastic. In the latter case, episodes of maser emission growth are called `flares'. The maser flares can last from several days (Fujisawa et al., 2012) to several years (MacLeod et al., 2018; Fontani, 2010). Fluctuations of maser emissions are attributed to corresponding changes in physical conditions in MYSOs, but specific mechanisms of maser flares are still a subject of debate. 

Different models of variability have been proposed (Fischer et al., 2023). Since the methanol masers are assumed to originate in the circumstellar disks of massive young stars, the variability is usually attributed to the processes in the disks. Variability of the order of years is attributed to the corresponding changes in the inner region of the circumstellar accretion disc. Various instabilities in the disk may cause the growth of the accretion rate and, as a consequence, the accretion luminosity of the star, which in turn affects the maser emission flux.  A widely accepted mechanism of this type is an episodic accretion from the gravitationally unstable disk (Vorobyov \& Basu, 2005; Caratti o Garatti et al., 2017; Meyer et al., 2017). Short time flares with durations of the order of days or weeks are probably related to dynamical processes  localized with the stellar surface, the near-star environment, or the circumstellar disk. In particular, (Chen et al., 2025) and (Khaibrakhmanov et al., 2025) proposed that the explosive energy release during magnetic reconnection near the stellar magnetosphere makes a significant contribution to the MYSO's luminosity and, correspondingly causes the maser flares. According to (Khaibrakhmanov et al., 2025), a current sheet may exist in the region between the disk and a star. The current sheet can exist in two states. In the quiescent state, magnetic reconnection is slow, which does not result in significant energy release. In this regime, the energy required for maser excitation does not reach a critical level. However, in the flare state, there is a sharp increase in the reconnection rate of the magnetic force lines. This rapid reconnection leads to a significant release of energy, thus leading to increase in the radiation flux pumping the maser. Such a variable radiation flux from the star causes the variability in the maser emission.

In this work, we further test the hypothesis of (Khaibrakhmanov et al., 2025) in application to several MYSOs, for which magnetic fields and maser flares are observed. Since we work within the framework of hypotheses, the selection criterion for target objects was based on published data confirming the presence of magnetic fields in maser-emitting regions. We compile observational data on masers in the NGC6334I, G33.641-0.228, and G12.89+0.49 regions, for which there are well defined parameters of the maser flares, as well as robust measurements of the magnetic field strength. Our aim is to test the hypothesis on the connection between magnetic reconnection and maser flares in application to example objects. Further, we plan to extend the sample and use our approach for a more complete study including other objects. To analyse the role of magnetohydrodynamical (MHD) effects in the considered objects, we estimate the stellar magnetospheric radius and compare stellar and disk's magnetic field strengths in this region.

 \section*{Objects sample}

In order to analyse the hypothesis of MHD nature of maser flares, we compile observational data on the objects, for which magnetic field measurements are available.
The compiled data for NGC6334I, G33.641-0.228, G12.89+0.49 are listed in Table~\ref{tab1}. The table contains: the name of the region (column~1); the distance from the Sun to the object in kpc (column 2); luminosity of the star in units of $\cdot10^{4} $$L_\odot$ (column 3); duration of the burst in days (column 4); magnetic field strength in mG (column 5). Data on G33.641-0.228 are given for comparison, since this object has already been analysed by )Khaibrakhmanov et al., 2025).

\begin{table*} [h]
\begin{center}
\begin{threeparttable}
\centering
\scriptsize
  \caption{Observational characteristics of objects}\label{tab1}
\begin{tabular}{l r r r r r r r r r }
\hline
    
 Name& Distance from the Sun  & Luminosity         & Burst   & Magnetic field \\
     & to object,        & of star,
     & duration, & strength, \\
   & kpc &  $\cdot10^{4} L_\odot$ & d & mG\\
   \hline
 NGC6334I & $1.74$  & $1.7$  & $385$ & $6$ $\pm 2$ \\

 G33.641-0.228 & $4$  & $12$ & $12$ & $$-18$$ $\pm 4$\\

 G12.89+0.49& $3.6$ & $3.1622$ & $150$ & $(4-12)$ $\pm 2$\\
 
\hline 
\end{tabular}
\end{threeparttable}
\end{center}
\end{table*}

Table data are collected from Bachiller and Cernicharo (1990), Kojima et al. (2018), Goedhart et al. (2009), MacLeod et al. (2018), Vlemmings, (2008), Wu et al. (2023), Vorster et al. (2024).

Table~\ref{tab1} shows that the duration of flares observed in the indicated objects varies from a few days to one year. The shortest flare among the considered objects is detected in  G33.641-0.228, the longest flare~--- in NGC6334I. 
Measurements of the Zeeman splitting indicate that the magnetic field strength detected in the masers ranges from $5$~mG (NGC6334I) to $18$~mG (G33.641-0.228). Following (Khaibrakhmanov et al., 2025), we assume that these measurements trace large-scale magnetic fields in circumstellar disks around massive stars in considered MYSOs. Generally speaking, young massive stars in the considered object should also posses a magnetic field. The interaction of the accretion flow from a circumstellar disk with stellar magnetic field leads to the formation of a magnetosphere around a star. In the following, we analyse this interaction and its possible connection to maser flares.

\section*{Analysis of the observational data}

In this section, we estimate the properties of gas in the masing regions, as well as analyse the magnetic fields in the vicinity of massive young stars from the sample.

Let us first estimate the distance from the star to the maser. We assume that the temperature of the accretion disk is determined by the heating caused by absorption of the star's radiation (Armitage P.J., 2019). The incident angle of stellar flux onto the disk's surface has a typical value $f=0.05$ (Akimkin et al., 2012). Asuming that the disk radiates as a blackbody, we can estimate the disk temperature $T_{\rm disk}$ from the equality between the stellar radiation flux, $L_\star/4\pi r^2\cdot f$, and disk radiation flux, $\sigma_{\rm sb}T_{\rm disk}^4$, where $\sigma_{\rm sb}$ is the Stephan-Boltzmann constant. Adopting typical temperature in a masing region $T_{\rm disk} = T_{\rm M}=100$~K, one estimate the distance to the maser:

\begin{center}
    \begin{equation}
        R_{M} = 380 \left(\frac{f}{0.05} \cdot \frac{L_{\star}}{1.2\cdot10^{4} L_{\odot}}\right)^{1/2} \left(\frac{T_{\rm M}}{100\,\mathrm{K} }\right)^{-2}.
     \end{equation}
  \end{center}
Using the observational data from Table~1, we derive that corresponding distances for considered objects range from $380$~au (G33.641-0.228) to $620$~au (G12.89+0.49). These values agree with the observational constraints by Sanna et al. (2017) and theoretical prediction of Guadarrama et al. (2024) regarding masing regions in circumstellar disks. 

Data on magnetic field strength in masing region (column~5 in Table~1) allow us to estimate the plasma beta, which is the ratio of plasma pressure to magnetic pressure:
\begin{center}
    \begin{equation}
        {\beta} = 0.01 \left(\frac{n_{m}}{10^{6} \text{cm}^{-3}} \right) \left(\frac{T_{m}}{100\,\mathrm{K}}\right)
        \left(\frac{B_{m}}{18\,\mathrm{mG}}\right)^{-2} \label{eq:beta}
     \end{equation}
  \end{center}
Equation~(\ref{eq:beta}) shows that plasma beta in considered regions range from $0.01$ (G33.641-0.228 ) to $0.09$ (NGC6334I). Such a small values of plasma beta indicates that the magnetic pressure dominates over the plasma pressure in the masing regions. Therefore, the magnetic field is strong and can affect the dynamics of the circumstellar disks.

The measurements of the Zeeman splitting in maser lines in cosidered regions indicate that the circumstellar disks around young massive stars have magnetic fields. The star formation theory predicts that both stars and their circumstellar disks are born with a fossil large-scale magnetic field originating from the magnetic field of parent molecular cloud (Dudorov \& Khaibrakhmanov 2015). Based on such considerations, we assume that not only disks, but also massive young stars in considered MYSOs have intense magnetic fields. The configuration of of stellar magnetic field cannot be studied observationaly at the moment. The magnetic field of low-mass young stars has been shown to be mainly dipole-like (Johnstone et al., 2013). In a first approximation, we use an analogy with low-mass young stars and suppose that the magnetic field of massive young stars has also dipole configuration. The interaction of such a magnetic configuration with an accreting matter leads to the formation of a magnetosphere around a star. The radius of the star's magnetosphere $R_{m}$ can be estimated from the equality between the viscous stresses in the disc and the Maxwell's stresses of the stellar magnetic field, 
  \begin{center}
    \begin{equation}
        \frac{R_{m}}{9R_{\odot}} = 1.1 \left(\frac{R_{s}}{9R_{\odot}} \right)^{12/7} \left(\frac{B_{s}}{1 \: \mbox{kG}} \right)^{4/7} \left(\frac{\dot{M}}{10^{-5}M_{\odot}/\mbox{yr}} \right)^{-2/7} \left(\frac{M_{s}}{10M_{\odot}} \right)^{-1/7}, \label{eq:Rmag}
     \end{equation}
  \end{center}
where we adopted typical radius of the star $9\,R_\odot$, corresponding to a spherical black body with luminosity 1.2 $\cdot10^{4} L_{\odot}$ and effective temperature $2\cdot10^4$ K.

Equation~(\ref{eq:Rmag}) shows that the magnetosphere radius may range from $\gtrsim 1$ to $13$ stellar radii. The radius increases with increasing magnetic field strength, decreases with increasing stellar mass. It also increases with decreasing accretion rate, since a smaller accretion rate corresponds to smaller ram pressure acting on the magnetosphere.

Assuming that the stars have dipole magnetic field configuration, we estimate the strength of the star magnetic field at the magnetosphere boundary:
\begin{center}
    \begin{equation}
        B_{star} = 750 \left(\frac{M_{\star}}{10M_{\odot}} \right)^{3/7} \left(\frac{B_{\star}}{1 \: \mbox{kG}} \right)^{-5/7} \left(\frac{\dot{M}}{10^{-5}M_{\odot}/\mbox{yr}} \right)^{6/7} \left(\frac{R_{\star}}{9R_{\odot}} \right)^{36/7}  \label{eq:Bstar_mag}
     \end{equation}
  \end{center}
Equation~(\ref{eq:Bstar_mag}) shows that the strength of the stellar magnetic field at the magnetosphere boundary range from $750$~G in G33.641-0.228 to $1095$~G in G12.89+0.49.

The magnetic field measurements (column 4 in Table~\ref{tab1}) can be used to estimate the strength of the stellar magnetic field at the magnetosphere boundary.
Based on the assumption that the protostellar discs in the considered objects are magnetostatic, the strength of the stellar magnetic field at the magnetosphere boundary can be estimated as $B_{disk}\propto n_{disk}^{1/2}$, where $n_{disk}$ is the disk's density (see Equation (A.11) in Khaibrakhmanov, 2025). We use the model of Shakura and Sunyaev (1973) to estimate $n_{disk}$ in application to considered objects.

The characteristics of magnetic fields in considered object are listed in Table~\ref{tab2}: the name of the region (column 1); radius of star in units of $R_{\odot}$ (column 2); magnetospheric radius in units of $R_{\star}$ (column 3); distance from the star to the maser in au (column 4); strength of disk's magnetic field at the magnetosphere boundary (column 5); strength of stellar magnetic field at the magnetosphere boundary (column 6); plasma beta (column 7).
 
\begin{table*} [h]
\begin{center}
\begin{threeparttable} 
\scriptsize
  \centering
  \caption{Theoretical estimations of magnetic fields in HMYSOs}\label{tab2}
\begin{tabular}{l r r r r r r r r r }
\hline
 Name  & $R_{\star}$, & $R_{m}$, & Distance,  &  $B_{disk}$, &  $B_{star}$, & $\beta$\\
 & $R_{\odot}$ & $R_{\star}$ & a.u. & Gs & Gs & \\
\hline
 NGC6334I & 11 & $3$~---$11$ & $454$ & $1170$& $890$ & $0.09$\\

 G33.641-0.228 & $9$ &$2$~---$9$ & $380$ & $1880$& $750$ & $0.01$\\

 G12.89+0.49& $14$&  $2$~----$8$  & $620$ & $590$ & $1095$ & $0.02$\\

 \hline 
\end{tabular}
\end{threeparttable}
\end{center}
\end{table*}

Table~2 shows that the strength of the disk's magnetic field in the region of the magnetospheric radius lie in the range from $590$ to $1880$~G. These values are comparable to the stellar magnetic field strengths (column 5 in Table~\ref{tab2}). Therefore, we can make an assumption that the magnetic field plays an important role in the dynamics of considered regions. In particular, the maser flares in NGC6334I and G12.89+0.49 can be related to magnetic reconnection events near stellar magnetospheres, as it was hypothesized by (Chen et al., 2025) in application to G36.11+0.55 and by (Khaibrakhmanov et al., 2025) in application to G33.641-0.228.

\section*{Discussion and conclusion}

We collected observational data on masers located in massive star formation regions such as NGC6334I, G33.641-0.228, and G12.89+0.49. In our analyses, we estimated the distance from the star to the maser, the magnetospheric radius, and the magnetic field strength near this radius. The values of the magnetospheric radius vary between $\gtrsim 1$ and 12 stellar radii, which depends on stellar and disk parameters. The estimated distance from the star to the masing regions is of $454$~au for NGC6334I, $380$~au for G33.641-0.228, and $620$~au for G12.89+0.49. The magnetic field strength near the magnetosphere of young stars has also been estimated. This value is $1170$~G for the NGC6334I region, $1880$~G for G33.641-0.228, and $590$~G for G12.89+0.49. In the considered objects, the stellar and disk's magnetic field strengths in the region of the magnetosphere are comparable.

 The estimates made in this work confirm the hypothesis stated by (Khaibrakhmanov et al., 2025) that luminosity bursts and corresponding maser flares can be of MHD nature. In particular, the current sheets may be present at the boundary of the magnetospheres. Fast magnetic reconnection in these sheets may contribute to overall luminosity of the MYSOs and cause maser flares. Our study shows that the hypothesis about the relation of the magnetospheric magnetic reconnection with maser variabillity is applicable to considered objects. This result allows us to conclude that the magnetic reconnection-related maser variability can be a common phenomenon in MYSOs. To test this conclusion, we aim to compile and analyse data on maser variability and magnetic fields in MYSOs for a larger sample of objects.

{\bf Acknowledgements.} The work of S. Khaibrakhmanov and A. Sobolev is supported by the Russian Science Foundation (project 23-12-00258).

\subsection*{\rm \bf \normalsize References}

\setlength\parindent{-24pt}

\par

Akimkin V. V., Pavlyuchenkov Ya. N., Launhardt R., Bourke T. (2012) Structure of CB 26 protoplanetary disk derived from millimeter dust continuum maps. Astronomy Reports 56, 915-930

Armitage P.J. (2019) Physical Processes in Protoplanetary Disks. Saas-Fee Advanced Course 45, 1

Ashimbaeva N.T., Lekht E.E., Krasnov V.V., Shoutenkov V.R. (2024) Investigation of H2O and OH masers in the region of formation of a young high-mass stellar object (S255 NIRS 3). Astronomy Reports 68, 453-471

Bachiller, R., Cernicharo, J. (1990) A\&A 239, 276

Beuther, H., Kuiper, R., \& Tafalla, M. (2025). Star formation from low to high mass: A comparative view. arXiv e-prints

Cragg D. M., Sobolev A. M., Godfrey P. D. (2002) MNRAS 331(2), 521-536

Caratti o Garatti, A., Stecklum, B., Garcia Lopez, R., Eislöffel, J., Ray, T. P., Sanna, A., Cesaroni, R., Walmsley, C. M., Oudmaijer, R. D., de Wit, W. J., Moscadelli, L., Greiner, J., Krabbe, A., \& Fischer, C. (2017). Disk-mediated accretion burst in a high-mass young stellar object. Nature Physics, 13(3), 276-279

Chen, X., Liu, J. T., Bayandina, O. S., et al. (2025). Magnetic reconnection as the driving factor behind high-mass protostellar luminosity flares. Communications Physics, 8, 249

Dudorov, A. E., \& Khaibrakhmanov, S. A. (2015). Theory of fossil magnetic field. Advances in Space Research, 55(3), 843-850

Elitzur M. (1992) "Astronomical masers." Annual Review of Astronomy and Astrophysics, 30, 75-112

Ellingsen, S. P., Voronkov, M. A., Cragg, D. M., Sobolev, A. M., Breen, S. L., \& Godfrey, P. D. (2007). Investigating high-mass star formation through maser surveys. In J. M. Chapman \& W. A. Baan (Eds.), Astrophysical Masers and their Environments (Vol. 242, pp. 213-217)

Fontani F., Cesaroni R., Furuya R. S. (2010) Class I and Class II methanol masers in high-mass star-forming regions. A\&A 517, A56

Fujisawa, K., Sugiyama, K., Aoki, N., Hirota, T., Mochizuki, N., Doi, A., Honma, M., Kobayashi, H., Kawaguchi, N., Ogawa, H., Omodaka, T., \& Yonekura, Y. (2012). Bursting activity in a high-mass star-forming region G33.640.21 observed with the 6.7 GHz methanol maser. Publications of the Astronomical Society of Japan, 64(1), 17

Fischer, W.~J., Hillenbrand, L.~A., Herczeg, G.~J., Johnstone, D., Kospal, A., \& Dunham, M.~M. (2023). Accretion variability as a guide to stellar mass assembly. In S. Inutsuka, Y. Aikawa, T. Muto, K. Tomida, \& M. Tamura (Eds.), Protostars and Planets VII (Vol. 534, p. 355). Astronomical Society of the Pacific Conference Series

Goedhart, S., Langa, M. C., Gaylard, M. J., Van Der Walt, D. J. (2009) MNRAS 398(2), 995-1010

Guadarrama R., Vorobyov E. I., Rab Ch., Güdel M., Caratti o Garatti A., Sobolev A. M. (2024) The influence of accretion bursts on methanol and water in massive young stellar objects. A\&A 684, A51

Hafner A., Green J.A., Burdon A., Popova E., Ladeyschikov D., Breen S., Burns R.A., Chibueze J.O., Gray M.D., Kramer B.H., MacLeod G., Sobolev A., Voronkov M. (2024) M2P2 I: Maser Monitoring Parkes Program data description and Stokes-I OH maser variability. PASA 41, e009

Johnstone C.P., Jardine M.M., Gregory S.G., Donati J.-F., Hussain G. (2014) Classical T Tauri stars: magnetic fields, coronae and star–disc interactions. Monthly Notices of the Royal Astronomical Society 437(4), 3202-3220

Khaibrakhmanov, S., Dudorov, A., \& Sobolev, A. (2018). "Dynamics of magnetic flux tubes and IR-variability of young stellar objects." Research in Astronomy and Astrophysics, 18(8), 090

Kojima, Y., Fujisawa, K., Motogi, K. (2018) IAU Symposium 336, 336-337

Khaibrakhmanov S. A. (2024) Magnetic fields of protoplanetary disks. A\&ATr 34, 139-162

Khaibrakhmanov, S. A., Sobolev, A. M., \& Chen, X. (2025). The magnetospheric magnetic reconnection in high-mass young stellar objects and its possible relation to methanol maser flares. Accepted to Astronomy \& Astrophysics.

Menten K. M. (1993) Astrophysical Masers, vol. 412, pp. 199-202

Meyer, D. M.-A., Vorobyov, E. I., Kuiper, R., \& Kley, W. (2017). On the existence of accretion-driven bursts in massive star formation. Monthly Notices of the Royal Astronomical Society, 464, L90–L94

MacLeod, G. C., Smits, D. P., Goedhart, S., Hunter, T. R., Brogan, C. L., Chibueze, J. O., van den Heever, S. P., Thesner, C. J., Banda, P. J., Paulsen, J. D. (2018) MNRAS 478(1), 1077-1092

Shakura, N. I., \& Sunyaev, R. A. (1973). Black holes in binary systems. Observational appearance. Astronomy \& Astrophysics, 24, 337-355

Sanna A., Moscadelli L., Surcis G., van Langevelde H. J., Torstensson K. J. E., Sobolev A. M. (2017) Planar infall of CH$_{3}$OH gas around Cepheus A HW2. A\&A 603, A94

Shakhvorostova N.N., Vol'vach L.N., Vol'vach A.E., Dmitrotsa A.I., Bayandina O.S., Val'tts I.E., Alakoz A.V., Ashimbaeva N.T., Rudnitskii G.M. (2018) Search for H2O maser flares in regions of formation of massive stars. Astronomy Reports 62, 584-608

Vorobyov, E. I., \& Basu, Shantanu. (2005). The Origin of Episodic Accretion Bursts in the Early Stages of Star Formation. The Astrophysical Journal Letters, 633(2), L137-L140

Vlemmings, W. H. T. (2008) A new probe of magnetic fields during high-mass star formation. Zeeman splitting of 6.7 GHz methanol masers, vol. 484, pp. 773-781

Vorster, J. M., Chibueze, J. O., Hirota, T., MacLeod, G. C., van der Walt, D. J., Vorobyov, E. I., Sobolev, A. M., \& Juvela, M. (2024). "Identifying the mechanisms of water maser variability during the accretion burst in NGC6334I." A\&A, 691, A157

Wu, Jiong-Heng, Chen, Xi, Zhang, Yan-Kun, Ellingsen, Simon P., Sobolev, Andrej M., Zhao, Zhang, Song, Shi-Ming, Shen, Zhi-Qiang, Li, Bin, Xia, Bo, Zhao, Rong-Bin, Wang, Jing-Qing, \& Wu, Ya-Jun. (2023). Physical environments of the luminosity outburst source NGC 6334I traced by thermal and maser lines of multiple molecules. The Astrophysical Journal Supplement Series, 265(2), 49

\end{document}